\documentclass[prb,twocolumn,showpacs,amsmath,amssymb,aps]{revtex4}

\usepackage{graphicx}%
\usepackage{dcolumn}
\usepackage{amsmath}

\begin{document}


\title[Short Title]{Local Hall effect in hybrid ferromagnetic/semiconductor devices}

\author{Jinki Hong}
\altaffiliation[Also at ]{Nano Devices Research center, Korea Institute of Science and Technology}
\author{Sungjung Joo, Tae-Suk Kim}

\author{Kungwon Rhie}
\email{krhie@korea.ac.kr}
\altaffiliation[Also at ]{Nano Devices Research center, Korea Institute of Science and Technology}

\author{K. H. Kim}
\author{S. U. Kim}
\affiliation{Department of Physics, Korea University, Chochiwon 339-700, Korea}

\author{ B. C. Lee}
\affiliation{Department of Physics, Inha University, Incheon 402-751, Korea}

\author{Kyung-Ho Shin}
\email{kshin@kist.re.kr}
\affiliation{Nano Devices Research center, Korea Institute of Science and Technology, Seoul 130-650,
Korea}

\date{\today}

\begin{abstract}
We have investigated the magnetoresistance of
ferromagnet-semiconductor devices in an InAs two-dimensional electron
gas system in which the magnetic field has a sinusoidal profile. 
The magnetoresistance of our device is large. The longitudinal resistance has an additional
contribution which is odd in applied magnetic field. It becomes even
negative at low temperature where the transport is ballistic. 
Based on the numerical analysis, we confirmed that our data can be
explained in terms of the local Hall effect due to the profile
of negative and positive field regions. This device may be useful for
future spintronic applications.
\end{abstract}



\maketitle

Recently, two-dimensional electron gas (2DEG) systems combined with micromagnets 
have attracted much attention not only for fundamental physics,\cite{R1} 
but also for device applications.\cite{R0,R2,R3,R5,SJ}
In hybrid ferromagnet-semiconductor devices, inhomogeneous magnetic field
is generated by micromagnets on top of 2DEG
and acts as the magnetic barrier to moving charge carriers. 
The complex charge build-up in the Hall bar leads to 
interesting phenomena in longitudinal as well as transverse resistance.   
This local Hall effect can be used as spintronic devices,\cite{R0} 
magnetic field sensors,\cite{R2}
non-volatile memory cells,\cite{R3} and micromagnetometers.\cite{R5}  
In this paper, we present the hybrid ferromagnet-semiconductor devices 
in which the longitudinal resistance is a mixture of even and odd contributions
in external magnetic field, but the transverse resistance is odd. 
This interesting feature of our device can be understood in terms of local Hall effect
caused by a sinusoidal magnetic field profile.

\begin{figure}[b!]
\includegraphics[width=5cm]{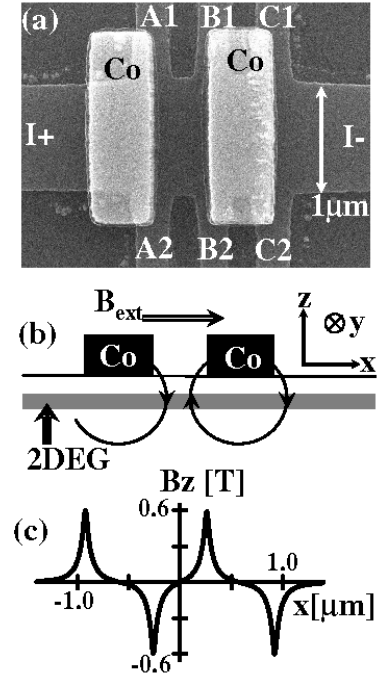}
\caption{(a) Scanning electron micrograph of the sample. 
(b) Schematic cross sectional view of the device structure. 
(c) Profile of $B_{z}$ along the $x$ axis on the 2DEG plane when the magnetization 
of the micromagnets (Co) is 1.8 T.}
\label{fig1}
\end{figure}

Our device structure is displayed in Fig.~1(a) as a scanning electron microscope image. 
An InAs 2DEG mesa, with a current channel (from $I+$ to $I-$) and six voltage probes 
(denoted as A1, B1, C1, A2, B2, and C2), was patterned by e-beam lithography.
The current channel is 1 $\mu$m wide and the voltage probes are separated about 0.6 $\mu$m 
from center to center. 
Two 300~nm thick micromagnets were deposited on top of the InAs 2DEG mesa by e-beam 
evaporation of Co. The InAs 2DEG resides 35.5 nm below the surface. 
Fig.~1(b) is a schematic cross sectional view of the sample and the profile of the fringe field 
generated near the edges of the micromagnets. The $x$-$y$ plane is defined by the 2DEG mesa 
with the $x$-axis along the current direction ($I+$ to $I-$). 
External magnetic field ($B_{\rm ext}$) aligns the magnetization ($M_x$) of micromagnets 
along the $x$-axis and the micromagnets generate the fringe field component ($B_z$) 
normal to the 2DEG plane. 
Calculated $B_z$ is plotted in Fig.~1(c) as a function of $x$ when micromagnets are 
fully magnetized along the $x$ direction. 
$B_z$ is sinusoidal with two maxima and two minima and is proportional to $M_x$. 
The centers of the voltage probes are positioned close to the maximum or minimum $B_z$. 
The mobility and the carrier density of the devices are 7.9~m$^2$/Vsec and $2.1\times
10^{16}$~m$^{-2}$, respectively, at 2~K and 2.0~m$^{2}$/Vsec and 
$1.9 \times  10^{16}$~m$^{-2}$ at 300 K.
 Resistance was measured using a standard lock-in technique with low-frequency (98~Hz) 
ac current of 1~$\mu$A.

\begin{figure}[t!]
\includegraphics[width=8.5cm]{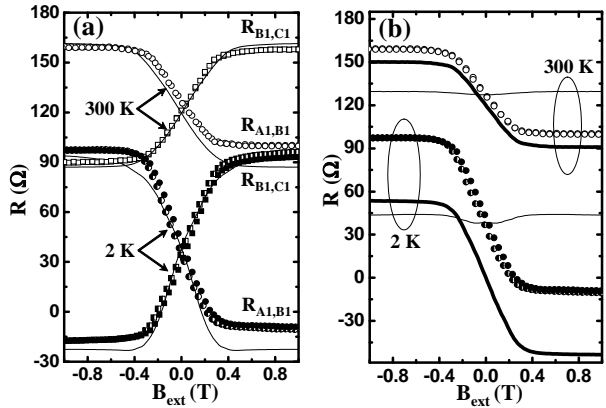}
\caption{ (a) Longitudinal resistance vs. external magnetic field $B_{\rm ext}$. 
The circles and rectangles  denote $R_{\rm A1,B1}$ and $R_{\rm B1,C1}$, respectively. 
The solid lines are calculated results (see the text).     
(b)$R_{\rm A1,B1}$ is decomposed into symmetric (thin solid lines) 
and anti-symmetric part (thick solid lines) in $B_{\rm ext}$. 
The anti-symmetric curve at 300 K is offset by 120 $\Omega$. 
At 2 K, filled symbols: $B_{\rm ext} \uparrow$
and half-filled symbols: $B_{\rm ext} \downarrow$. The hysteresis is
almost indistinguishable for empty symbols at 300 K.
 } \label{fig2}
\end{figure}

In Fig.~2(a), longitudinal resistances $R_{\rm A1,B1}$ (between A1 and B1)
and $R_{\rm B1,C1}$ are plotted as a function of external magnetic field at 2~K and 300 K.
Two curves of $R_{\rm A1,B1}$ and $R_{\rm B1,C1}$ are almost symmetric each other with respect 
to the resistance axis.
$R_{\rm A2,B2}$ ($R_{\rm B2,C2}$) was also measured and was nearly the same 
as $R_{\rm A1,B1}$ ($R_{\rm B1,C1}$).
$\Delta R$, defined by the difference between the maximum and the minimum longitudinal resistance, 
is relatively large.
$\Delta R$ is  60~$\Omega$ at 300~K and 107~$\Omega$ at 2~K for $R_{\rm A1,B1}$.
The sensitivity of longitudinal resistance to $B_{\rm ext}$ is estimated with 
$dR/dB_{\rm ext}$ at zero field, whose value is $113 \Omega$/T at 300~K and 178~$\Omega$/T at 2~K.
In a uniform  magnetic field normal to the 2DEG system,
the longitudinal resistance (magnetoresistance) is an even function of the external 
magnetic field while the transverse (Hall) resistance is an odd function. 
However, the longitudinal resistance in our device has both even (magnetoresistance) and 
odd (Hall) contributions as displayed in Fig.~2(b), and becomes even {\it negative} in high field at 2 K.
This interesting feature in our longitudinal resistance in fact arises from 
local Hall effect due to a sinusoidal variation of local magnetic field, as we shall explain below.

Fig.~3 shows the transverse resistance, $R_{\rm A2,A1}$ and $R_{\rm B2,B1}$.
$R_{\rm C2,C1}$ (not shown) is almost the same as $R_{\rm A2,A1}$.
Taking into account a possible minor misalignment of Hall voltage probes, 
we may conclude that the transverse resistance is an odd function of external magnetic field,
as in a uniform magnetic field. 
The approximate relation, $R_{\rm A2,A1} (B_{\rm ext}) \approx R_{\rm B2,B1}(-B_{\rm ext})$,
means that $B_z$ near A1 and A2 is directed opposite to $B_z$ near B1 and B2, 
but the field strength is almost equal at both regions. 
Our data clearly support the sinusoidal variation of local magnetic field on the 2DEG system.

\begin{figure}[t!]
\includegraphics[width=6.5cm]{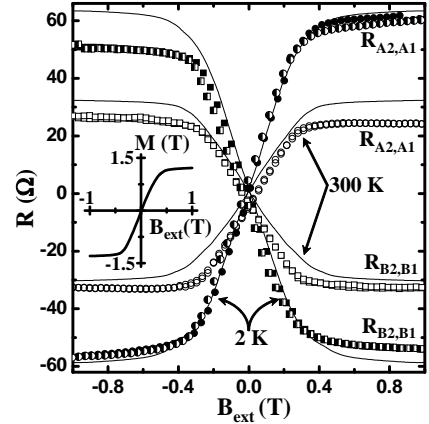}
\caption{ Transverse resistance vs. $B_{\rm ext}$. 
The circles are for $R_{\rm A2,A1}$ and the rectangles
are for $R_{\rm B2,B1}$. The filled and half-filled symbols have the same meaning as
in Fig. \ref{fig2}(a). The inset is a functional relation between the
magnetization of Co and $B_{\rm ext}$, which was found from fitting.} \label{fig3}
\end{figure}

In order to elucidate the origin of our interesting experimental data, we calculated 
the resistance numerically in the diffusive as well as ballistic transport regimes. 
The diffusive transport model \cite{R7} is adopted at 300 K (the mean free path of electrons, 
$\ell_{\rm mfp} =0.46 \mu$m at 300 K, is shorter than the device size.) 
The electrostatic potential and the current density were obtained by solving 
the continuity equation \cite{R7} with the spatial dependent conductivity tensor (inhomogeneous 
magnetic field). 
We adopted the ballistic Landauer-B$\rm{\ddot{u}}$ttiker formalism~\cite{LB1} at 2~K 
($\ell_{\rm mfp} = 1.9 \mu$m is longer than the device size.)
Transmission probability was computed \cite{particle1, particle2} by injecting electrons 
with Fermi velocity from one lead and by counting the number of electrons exiting the other lead.
In the meantime, injected electrons are subject to Newton's equation of motion 
\cite{particle1, particle2} under the magnetic barriers.

 In our numerical simulations, we took into account the device geometry including the shape 
of the voltage probes. 
The profile of local magnetic field ($B_z$) [see Fig.~1(c)] is obtained from dipole field \cite{dpfield}
of micromagnet magnetization $M_x$. 
After computing the resistance as a function of $M_x$, we tried to fit the experimental 
transverse resistance $R_{\rm A2,A1}$ with theoretical one (solid lines in Fig.~\ref{fig3}), 
by adjusting the $M_x$ (theory) versus $B_{\rm ext}$ (experiment) curve. 
The resulting functional form $M_x(B_{\rm ext})$ at 2 K (not shown for 300 K) 
is shown in the inset of Fig.~\ref{fig3} and is similar to the Brillouin function as expected. 
Other solid lines in Fig.~\ref{fig2}(a) and Fig.~\ref{fig3} are theoretical resistance curves with
adjusted $B_{\rm ext}(M_x)$.  
To conclude, we were able to reproduce the main features of our resistance curves 
in terms of local Hall effect due to a sinusoidal magnetic field [Fig.~1(c)].

\begin{figure}[t!]
\includegraphics[width=8.0cm]{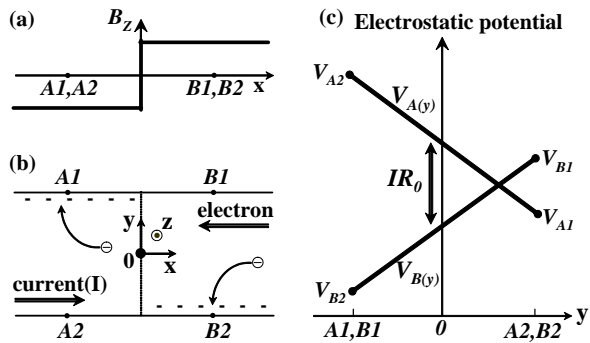}
\caption{(a) Model field profile of a step function.
(b) Schematics of electron trajectories and piling-up of charges. 
(c) Electrostatic potentials due to local Hall effect.}
\label{fig4}
\end{figure}

 In order to better understand the transport process in the Hall bar, 
we consider an inhomogeneous field distribution of Fig.~\ref{fig4}.
This field profile of a step function mimics the profile of $B_z$ 
in our device [see Fig.~1(c)] from A1 (negative minimum) to B1 (positive maximum).  
The schematic of the current channel is displayed in the $x$-$y$ plane in Fig.~\ref{fig4}(b).
Moving electrons are deflected by the Lorentz force under magnetic field ($B_z$) and piled up 
on the side of the device.
Because $B_z$ is positive (negative) at the right (left) side, electrons are piled up on the B2 (A1) 
side as displayed.
This local Hall effect explains our resistance data as we shall show below.

 The accumulation of negative charges lowers the electrostatic potential
at A1 (B2) with respect to A2 (B1).
The resulting electrostatic potentials are displayed in Fig.~\ref{fig4}(c).
$V_{\rm A}(y)$ [$V_{\rm B}(y)$] is the potential along the straight line from A2 to A1 [B2 to B1] 
in the sample. 
When $B_z$ is zero, we have $V_{\rm B2}$=$V_{\rm B1}$ and $V_{\rm A2}$=$V_{\rm A1}$, 
and $V_{\rm A}(y)$ and $V_{\rm B}(y)$ will be simply constant functions. 
In this case, the voltage difference between two curves can be expressed as $IR_0$ 
from Ohm's law [$V_{\rm A}(y)-V_{\rm B}(y)=IR_0$], where $I$ is the current 
and $R_0$ is the longitudinal resistance in the absence of magnetic field. 
As magnetic field is introduced, charge piling-up gives rise to $V_{\rm A}(y)$ and $V_{\rm B}(y)$ 
in Fig.~\ref{fig4}(c). 
Due to the symmetry of charge distribution in Fig.~\ref{fig4}(b), the voltage difference 
at $y=0$ [$V_{\rm A}(y=0)-V_{\rm B}(y=0)$] in Fig.~\ref{fig4}(c) remains as $IR_0$, 
which is denoted by an arrow. 
Further the Hall voltages $V_{\rm B2,B1}=V_{\rm B2}-V_{\rm B1}$ and 
$V_{\rm A2,A1}=V_{\rm A2}-V_{\rm A1}$ have the same magnitude with an opposite sign, 
thanks to the symmetry in Fig.~\ref{fig4}.
Combining the above results, we can express $R_{\rm A1,B1}(B_{\rm ext})$ as
\begin{equation}
\label{RHR}
R_{\rm A1,B1}(B_{\rm ext}) = R_0 - R_{\rm A2,A1}(B_{\rm ext}),
\end{equation}
where $R_{\rm A2,A1}(B_{\rm ext})$ is the transverse (Hall) resistance between probes A2 and A1.
This relation explains why the longitudinal resistance curve has a term odd in $B_{\rm ext}$. 
Though Eq.~(\ref{RHR}) is derived from a simple field profile, it is quite amusing to note that
$-R_{\rm A2,A1}$ is comparable to antisymmetric part of $R_{\rm A1,B1}$ in Fig.~\ref{fig2}(b)
when $R_0$ is considered as symmetric part.

When electrons are piled up at A1 with respect to B1 sufficiently or $B_z$ is sufficiently 
strong in the right direction, locally induced Hall field between A1 and B1 can 
overcome applied current-driven electric field. 
This local inversion of the electric potential profile leads to the observed 
negative longitudinal resistance (NLR). 
We can also explain the NLR in terms of Eq.~(\ref{RHR}). 
The mobility increases with lowering temperature, which results in decrease of $R_0$ and 
increase of $R_{\rm A2,A1}(B_{\rm ext})$ (see Fig.~\ref{fig3}.) 
If $B_z$ is strong enough for $R_{\rm A2,A1}$ to be larger than $R_0$, the NLR occurs.

In summary, we fabricated ferromagnet-semiconductor hybrid devices in which the dipolar field 
of micromagnets is of a sinusoidal form on the Hall bar.  
The longitudinal resistance is featured with mixing of even and odd terms 
in external magnetic field ($B_{\rm ext}$), while the transverse (Hall) resistance is odd.  
Due to odd term in $B_{\rm ext}$, our longitudinal resistance becomes even negative
in high field at low temperature.
Based on numerical calculations in the diffusive and ballistic transport regimes, 
we confirmed that our interesting data can be explained in terms of 
the combined local Hall effect in the positive and negative magnetic field regions. 
We suggested a simple model which can explain the main results of our experiments.
Our result demonstrates that the control of local magnetic field by
micromagnets can provide another degree of freedom for device applications.

This work was supported by the Vison21 Program at Korea Institute of
Science and Technology, the SRC/ERC program of MOST/KOSEF (R11-2000-071), and
the Basic Research Program of KOSEF (R01-2006-000-11391-0).

\end{document}